\documentclass[conference]{IEEEtran}
\IEEEoverridecommandlockouts
\usepackage{cite}
\usepackage{amsmath,amssymb,amsfonts}
\usepackage{algorithmic}
\usepackage{hyperref}
\usepackage{graphicx}
\usepackage{textcomp}
\usepackage{xcolor}
\def\BibTeX{{\rm B\kern-.05em{\sc i\kern-.025em b}\kern-.08em
    T\kern-.1667em\lower.7ex\hbox{E}\kern-.125emX}}
\begin{document}

\title{Interactive Bayesian Generative Models for Abnormality Detection in Vehicular Networks}

\author{\IEEEauthorblockN{Nobel~J.~William\textsuperscript{}, Ali~Krayani\textsuperscript{},  Lucio~Marcenaro\textsuperscript{}, and Carlo~Regazzoni\textsuperscript{}}
\IEEEauthorblockA{\textsuperscript{}\textit{DITEN, University of Genoa, Italy} \\
emails:
nobel.johnwilliam@edu.unige.it, ali.krayani@ieee.org, \{lucio.marcenaro, carlo.regazzoni\}@unige.it\
}

}

\maketitle

\begin{abstract}
The following paper proposes a novel Vehicle-to-Everything (V2X) network abnormality detection scheme based on Bayesian generative models for enhanced network self-awareness functionality at the Base station (BS). In the learning phase, multi-modal data signals contrived by the vehicles’ integrated and sensing module are imbued into data-driven Generalized Dynamic Bayesian network (GDBN) models. Following that, during the testing phase, an Interactive Modified Markov Jump Particle filter (IM-MJPF) is utilized to forecast forthcoming network states and vehicle trajectories by leveraging the assimilated semantics embedded in the coupled multi-GDBNs. This approach involves learning statistically correlated association between evolving trajectories and network communication links. Security and surveillance of Internet of Vehicles (IOVs) links are performed online with high detection probabilities by matching predicted with observed network connectivity maps (graphs).
\end{abstract}


\section{Introduction}
Vehicular networks such as Vehicle-to-everything (V2X), consisting the vehicle-to-vehicle (V2V), vehicle-to-pedestrian (V2P), vehicle-to-infrastructure (V2I), and vehicle-to-network (V2N) communication are envisaged to be key components in Intelligent Transportation Systems (ITS)~\cite{10101}. 
Integrated Sensing and Communication (ISAC) is expected to play a pivotal role in the future V2X standard by enabling next-generation connected autonomous vehicles (CAVs) \cite{9919739}.
Moreover, the inherent openness and dynamism of V2X results in a restricted availability of precise Channel State Information (CSI). Consequently, ensuring secure transmission at the Physical Layer (PHY) poses a noteworthy challenge in ISAC-aided V2X systems. 
In order to address these real-world conditions, the implementation of ISAC technology necessitates the utilization of advanced machine learning techniques that effectively integrate communication and sensing data. This integration enables the development of ambient (situational) awareness and thus enabling cognitive services at the global network level, to systematically mitigate potential threats or detect abnormalities.
In high-mobility V2X scenarios, it is crucial for the transmitter or base station (BS) to possess the capability to anticipate connectivity requirements beforehand. This predictive ability allows for effective attack mitigation and resource allocation to ensure compliance with the demanding QoS requirements for V2X communications. 

Networked Sensing (NS) involves utilizing components of cellular networks, such as Base Stations (BS) or Roadside Units (RSUs), along with vehicular nodes to enhance the overall awareness of V2X networks \cite{wild2021joint}. This approach provides diverse perspectives on the communication and sensing environment. However, V2X networked sensing requires more attention to improve the estimation of network topology and enhance resource management and network configuration. Also, the integrity of V2I and V2V links is paramount for effective resource management in dense V2X environments, as jamming attacks can severely impact road vehicle safety and traffic efficiency. The V2I links include collaborative information showing the vehicle's location and other details from active V2V links exchanging basic safety messages (BSMs) and cooperative awareness messages (CAMs). This enables the receiving cognitive BS/RSU to learn a semantic mapping between vehicular trajectory states and their respective V2V network connectivity states. Unfortunately, there is limited research in this area. However, the Bayesian method can aid in inferring network structure even with unjustified measurement errors \cite{young2020bayesian}.
Furthermore, it is vital to anticipate how the network topology will change over time to ensure the safety of V2X operations by using the rules learned from past experiences. This will help in identifying potential issues like jamming or spoofing attacks that could cause abnormalities (i.e., network failure) and determine their root cause. By understanding the cause of the problem, effective measures can be taken to prevent such issues from happening again in the future.

%
%
Ensuring the safe operation of V2X applications requires detecting abnormalities in the topology caused by sensor attacks or new mobility motions, as well as network abnormalities caused by jamming or spoofing attacks. Several studies have addressed these concerns. For instance, \cite{10323101} proposes a framework for detecting and classifying sensor attacks that use GPS and LiDAR sensors. In \cite{9788452}, researchers analyze the impact of abnormal driver behaviour on data transmission in VANET networks. In \cite{8500715}, a new approach to driver profiling and abnormality detection is proposed, using LSTM for modelling drivers. Additionally, \cite{9484071} suggests a method for identifying jammers that affect vehicular networks. However, previous research has shown that multi-modal sensors, such as GPS, LiDAR, video cameras, and RF antennas, typically work independently to achieve separate goals and do not collaborate to improve V2X performance via abnormality detection, consequently enhancing security and resource management. 
Moreover, a PHY layer security oriented jammer detection scheme using Dynamic Bayesian network (DBN) was proposed in \cite{9825566} to detect attacks on V2I links, however, this work doesn't provide a paradigm for estimating V2X connectivity state at the global network level for abnormality detection. Another related work in \cite{russo2018anomaly} discusses the use of neural network-based Long Short-term Memory (LSTM) and Multilayer Perceptron (MLP) system for anomaly detection in vehicular communications. However, these deep learning techniques lack the ability to explain explicitly the causal probabilistic relationships that result in network abnormalities.

In this work, inspired by \cite{10118852}, we propose a new method of combining vehicle trajectory data collected by a video camera sensor with RF signals received from each vehicle in order to detect any abnormalities in the V2X network caused by jamming attacks.
%
%
This cognitive service is designed to identify network disruptions at the BS/RSU by combining multiple sensory data in a joint probabilistic representation. The method uses a probabilistic representation structured in a coupled Multi-Generative Dynamic Bayesian Networks (M-GDBN) generated during training. It also utilizes real-time vehicle trajectory measurements to predict the current and future V2X network connectivity structure. The Interactive Modified Markov Jump Particle filter (IM-MJPF) is used to make these predictions by inferring what the vehicles could transmit based on their mobility pattern. The BS/RSU performs the inference, considering the full or partial observability of both the vehicular trajectories and the V2X network's connectivity state/structure. The proposed method enables the BS/RSU to jointly predict the evolution of both vehicles' trajectories and RF signals and detect any abnormality causing a network failure at the high abstraction level in the probabilistic representation. It can also explain the cause behind the detected abnormalities and the sensor it is coming from (e.g., video camera sensor or RF antenna) through the causal probabilistic relationships encoded in the representation.
In this study, we conducted simulations to avoid the complexity of collecting data from multiple vehicles during the initial training phase. This would have been difficult to implement in real-world empirical analysis. We assumed that the Channel State Information (CSI) reported to the BS was accurate because of the multi-agent (vehicle) environment perception in our scenario. This integration enables network-level cognitive services in the V2X context, specifically aimed at mitigating anomalies. 

\section{System Model and  Problem Formulation}
\begin{figure}[t!]
    \centering
    \includegraphics[scale=0.25]{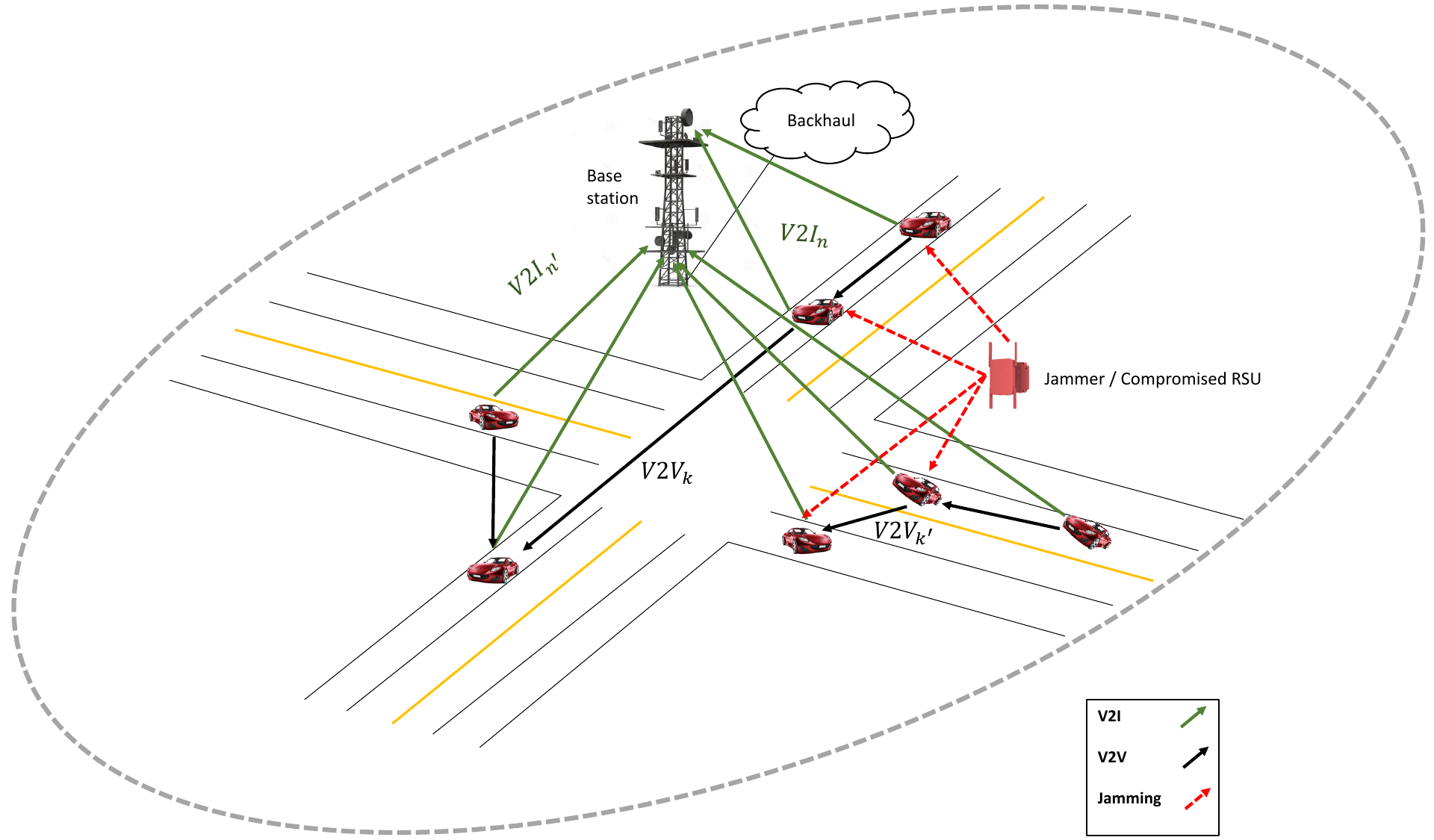}
    \caption{Illustration of the System model.}
    \label{fig_envir}
\end{figure}
Assume $\mathrm{N}$ connected vehicles traversing a freeway within a single-cell vehicular network, each transmitting it's respective positional information $p_{n,t} = [x_{n,t},y_{n,t}]$ where $\mathrm{n}\in\mathrm{N}$ at each time instant $\mathrm{t}$ jointly with supplementary information regarding it's active V2V connections and CSI (\textit{i.e.} channel quality, signal strength, interference, and other relevant metrics) to the BS located at $p_{BS} = [x_{BS},y_{BS}]$ via the $\mathrm{n^{th}}$ V2I link. Each vehicle exchanges messages with other vincinal vehicles $n'$ located at $p_{n',t} = [x_{n',t},y_{n',t}]$ on the freeway via the $\mathrm{k^{th}}$ V2V links. The total bandwidth available is divided into $\mathrm{N}$ physical resource blocks (PRBs) and each vehicle on the freeway accesses a single PRB  from the set of mutually orthogonal spaced frequency bands denoted as $\mathcal{F}=\{f_{1}, f_{2}, \dots, f_{N}\}$ for its V2I link (without loss of generality we assume $\mathrm{N} = \mathcal{F}$ here).
The set of V2V links $\mathcal{K}= \{f_{1}, f_{2}, \dots, f_{K}\}$ reuse the V2I up-link in an interference free optimal manner as discussed in~\cite{liang2018graph}. Moreover, scheduling and interference management of V2V traffic is assisted by the BS (in Mode 3) via the control signaling that assigns the resources used for V2V communications in a dynamic manner. It is to be duely noted in this scenario, the BS effectively assists in V2V links CSI management (not necessarily always done in mode 3). 
The antagonist here is a \textit{constant (single)} and \textit{periodic (multi-attack)} road side jammer/compromised RSU residing within the cell at $p_{j} = [x_{j},y_{j}]$. The system model is illustrated as in Fig.~\ref{fig_envir}. The intelligent jammer injects interference at mulitple frequencies (PRBs) and overloads the dynamic V2V links. 
The jamming detection problem (for each vehicle) can be devised as the following binary hypothesis test:
\begin{equation}\label{hypothesisTest}
\left\{\begin{array}{l}
\mathcal{H}_0: \mathrm{z}_{t, k}=\mathrm{g}_{t, k}^{n n'} \mathrm{x}_{t, k}+\mathrm{v}_{t, k}, \\
\mathcal{H}_1: \mathrm{z}_{t, k}=\mathrm{g}_{t, k}^{n n'} \mathrm{x}_{t, k}+\mathrm{g}_{t, k}^{j n'} \mathrm{x}_{t, k}^j+\mathrm{v}_{t, k}, \\
\end{array}\right.
\end{equation}
\noindent where $\mathcal{H}_0$ and $\mathcal{H}_1$ denote absence and presence of jammer, respectively. $\mathrm{x}_{t, k}$ is the reference signal, $\mathrm{v}_{t, k}$ is the additive white Gaussian noise (with variance $\sigma^{2}_n$) and $\mathrm{z}_{t, k}$ is the received signal at the individual vehicle transreceiver at time $\mathrm{t}$ over the $\mathrm{k^{th}}$ V2V link. $\mathrm{g}_{t, k}^{n n'}$ is the channel gain between vehicles $n$ and $n'$. $\mathrm{x}_{t, k}^j$ is the jammer signal that overwhelms the vehicle radio receivers with channel power gain $\mathrm{g}_{t, k}^{j n'}$ at $n'$ vehicle. $d^{nn'}_{t} = \sqrt{[(x_{n,t} - x_{n',t} )^2 + (y_{n,t} - y_{n',t} )^2}$ is the distance between vehicle $n$ and vehicle $n'$ at time $t$.

Mean while for the V2I links, the channel gain $g^{nBS}_{t,n}$ of the $n^{th}$ V2I link to the BS is $g^{nBS}_{t,n} = \alpha^{nBS}_{t,n} h^{nBS}_{t,n}$, where $\alpha^{nBS}_{t,n}$ = $\mathrm{G\beta} d^{-\gamma}_{t,nBS}$ captures the large-scale fading effects (path-loss and fading component). $\mathrm{G}$ is the path-loss constant and $\beta$ is log normal shadow fading random variable, $d^{nBS}_{t} = \sqrt{[(x_{n,t} - x_{BS} )^2 + (y_{n,t} - y_{BS} )^2}$ is the distance between $\mathrm{n^{th}}$ vehicle and the BS at time $t$. $\gamma$ is power decay exponent and $h_{t,n}$ is the small-scale fading component distributed according to $\mathcal{N}(0,1)$ as seen in \cite{guo2019resource}.   
%
To accommodate the global network perspective, we reformulate on road vehicles as \textit{nodes} and the communication links between them at time $t$ as \textit{edges}. We assume unweighted and undirected edges here to represent existence of bidirectional connection between vehicles $n$ and $n'$. Thus, a \textit{discrete time dynamic graph} $\mathcal{G}_{t} = ({V_t, \mathcal{E}_t})$ with $V$ vertices (nodes) and $\mathcal{E}$  edges as mapped in adjacency matrix $\mathbf{A} \in \mathbb{R}^{V \times V}$ where $V= N$. The binary Hypothesis test mentioned in \eqref{hypothesisTest} is modified to a graph perturbation scenario as follows:
\begin{equation}\label{GraphhypothesisTest}
\left\{\begin{array}{l}
\mathcal{H}_0: \mathrm{Z}_{\mathcal{G}_{t}}=\mathrm{G}_{t}+\mathrm{v}_{t} \\
\mathcal{H}_1: \mathrm{Z}_{\mathcal{G}_{t}}= \mathrm{f}(\mathrm{G}_{t})+\mathrm{v}_{t}  \\
\end{array}\right.
\end{equation}
where $\emph{f}(\mathrm{G}_{t})$ is a graph perturbation function informed by the jammer, $\mathrm{v}_{t}$ is the process noise and $\mathrm{Z}_{\mathcal{G}_{t}}$ is the observed graph.

\section{Proposed Scheme for Joint Network Structure Inference and Abnormality Detection}
\subsection{Environment Representation}
We represent the physical and RF environment (sensory signals) onto a \textit{model based data driven framework} that internally embodies a Generative Bayesian probabilistic model. Here, BS transceiver observes measurements $Z_t^m$ reported from individual vehicles (which includes position, estimated CSI and activated V2V connections) on the freeway at each time instant $t$. The individual physical states (positions) of $\mathrm{N}$ vehicles at time $t$ are mapped at a meta-cluster level termed as a Word $W^p_t$ within a learned global odometry/positional dictionary $D^p$. Each Word comprises of $N$ symbolised letters (aka clusters or superstates) $S^p_n$ with unique mean $\mu$ and variance $\Sigma$. Similarly, the activated V2V connections at each vehicle reported to the BS are contained in symbolized letter as $S^c_n$. It may be pertinent to note that several symbols fuse to make a letter here and is a function of the number of vehicles on the freeway. Multiple letters ($S^c_n$) based on data received at the BS imbibe into the Word $W^c_t$ representing the network connectivity state/map at time $t$. A global dictionary $D^c$ containing multitude of Words $W^c$ expressing the universe of learned vehicular network configurations is formulated during training.
The dynamic $m$-th data signal evolution function using Generalized State Space model at multiple hierarchical levels can be expressed as:
\begin{equation}\label{eq_discreteDynamic_model}
    \mathrm{\Tilde{W}^m}_{t} = \mathrm{{{f}^1}\hspace{1 pt}(\Tilde{W}^m_{t-1})}+ \Tilde{w}^m_{t},    
\end{equation}
\begin{equation}\label{eq_discrete2Dynamic_model}
    \mathrm{\Tilde{S}^m}_{t,n} = \mathrm{{{f}^2}\hspace{1 pt}(\Tilde{S}^m_{t-1,n})}+ \Tilde{w}^m_{t,n} ,
\end{equation}
where $m \in $ (\underline{p}ositional, \underline{c}ommunication) and $\mathrm{f}(.)$ is a non-linear function of the argument and $w^m_{t}$, $w^m_{t,n}$ encode the noise affecting the processes at different hierarchical levels. The discrete equations \eqref{eq_discreteDynamic_model} and \eqref{eq_discrete2Dynamic_model} illustrate the temporal evolution of Generalized Words and Generalized letters respectively during the higher latent state evolutionary process. \eqref{eq_discreteDynamic_model} maps transitions from previous Generalized Word to the current Word, embodying the dynamics of transitions between discrete-level positional or network configurations over time.
Moving down the hierarchical network, the continuous state dynamic evolution equation (expressed in \eqref{eq_contin1Dynamic_model}) and the measurement model mapping the continuous state to the observation space (in \eqref{eq_contin2Dynamic_model}) are defined as follows:
\begin{equation}\label{eq_contin1Dynamic_model}
    \mathrm{\Tilde{X}^m}_{t,n} = \mathrm{A\Tilde{X}^m_{t-1,n}}+ \mathrm{B\Tilde{U}}_{\Tilde{S}^{m}_{t,n}}+ \Tilde{w}^m_{t,n} ,
\end{equation}
\begin{equation}\label{eq_contin2Dynamic_model}
    \mathrm{\Tilde{Z}^m}_{t,n} = \mathrm{H\Tilde{X}^m_{t,n}}+ \Tilde{w}^m_{t,n} .
\end{equation}
More specifically, the motion dynamics of each individual $n^{th}$ vehicle in the cell or the V2V connections states are baked into \eqref{eq_contin1Dynamic_model}. To account for measurement uncertainty, the noise term $\Tilde{w}^m_{t,n}$ is introduced. $\mathrm{A} \in \mathbb{R}^{2d,2d}$ and $\mathrm{H} \in \mathbb{R}^{2d}$ are the dynamic and control matrices, respectively. $\mathrm{\Tilde{U}}_{\Tilde{S}^{m}_{t,n}}$ constitutes the control vector which represents the dynamic rules of the signal's temporal evolution. The measurement model in \eqref{eq_contin2Dynamic_model} encapsulates the mapping from the observed sensory signals $\mathrm{\Tilde{Z}^m}_{t,n} $
to the underlying hidden state $\mathrm{\Tilde{X}^m_{t,n}}$ through the measurement matrix 
$\mathrm{B} \in  \mathbb{R}^{2d}$ where $d$ is data dimensionality and depends on the features used  for training such as XY-coordinates, velocities, relative distances, RF IQ data, \textit{etc.}

\subsection{Learning Dictionaries (containing multi-GDBNs)}
\begin{figure}[!ht] 
    \centering
        \includegraphics[scale=0.265]{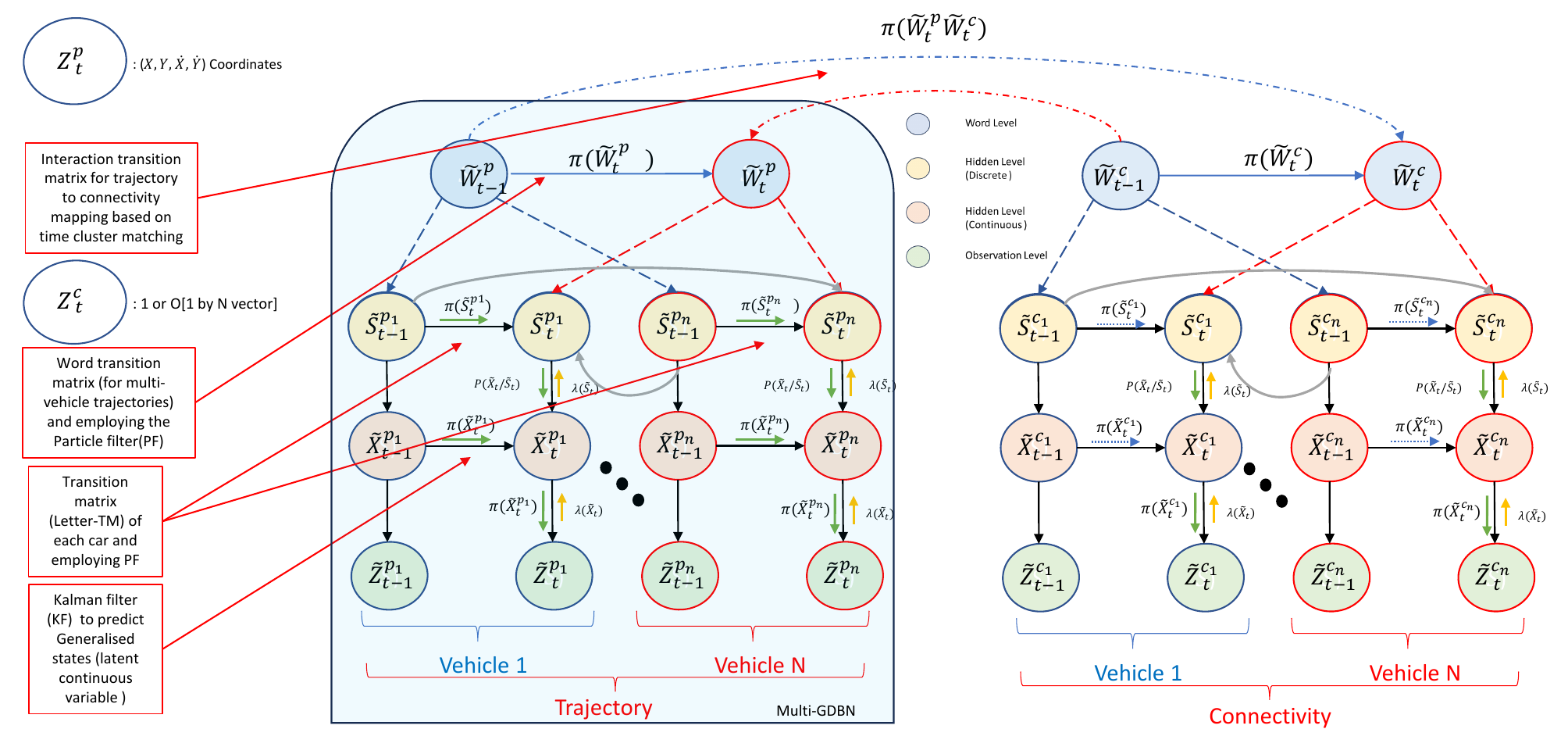}
    \caption{Cognitive Architecture: Interactive Coupled Multi-GDBN model.}
    \label{fig_InterCoupled_MultiGDBN}
\end{figure}

The dynamic hierarchical models expressed as \eqref{eq_discreteDynamic_model}, \eqref{eq_discrete2Dynamic_model}, \eqref{eq_contin1Dynamic_model} and \eqref{eq_contin2Dynamic_model}, provide us with the scaffold to build the Generalized Dynamic Bayesian Network \cite{9858012} (refer to Fig.~\ref{fig_InterCoupled_MultiGDBN} "blue-shaded region") that embodies conditional dependencies between latent variables and measured sensory states into a structure known as `probabilistic graphical model' (PGM). The generative process model imbibed into the PGM, explaining the sensory signals can be factorised as follows:
\begin{multline}\label{LearningDictionaries1}
\scriptsize
\mathrm{P}(\mathrm{W}_t^m, \mathrm{\Tilde{S}}^{m}_{t,n}, \mathrm{\Tilde{X}}^m_{t,n}, \mathrm{\Tilde{Z}}^m_{t,n}) =  
\mathrm{P}(\Tilde{W}^0_t) \mathrm{P}(\Tilde{S}^0_{t,n })
\mathrm{P}(\Tilde{X}^0_{t,n }) \\ \prod_{t}P(\mathrm{W}_t^m) \prod_{t,n}\mathrm{P}(\mathrm{\Tilde{S}}^{m}_{t,n}| \mathrm{W}_t^m) \mathrm{P}(\mathrm{\Tilde{X}}^{m}_{t,n}| \mathrm{\Tilde{S}}_{t,n}^m) \mathrm{P}(\mathrm{\Tilde{Z}}^m_{t,n}|\mathrm{\Tilde{X}}^m_{t,n}),
\end{multline} 
\noindent where $\mathrm{P}(\Tilde{W}^0_t)$, $\mathrm{P}(\Tilde{S}^0_{t,n })$ and $\mathrm{P}(\Tilde{X}^0_{t,n })$ are initial prior distributions. The  likelihood is $  
\mathrm{P} (\mathrm{\Tilde{Z}}^m_{t,n}|\mathrm{\Tilde{X}}^m_{t,n})$.  

\noindent The transition probabilities depicting the hierarchical and temporal relationships of the generalized state-space model are $\mathrm{P}(\mathrm{\Tilde{X}}^{m}_{t,n}| \mathrm{\Tilde{S}}_{t,n}^m)$ and $\mathrm{P}(\mathrm{\Tilde{S}}^{m}_{t,n}| \mathrm{W}_t^m)$.
The causality relationship chain between the random hidden and observed variables  $\mathrm{{W}_t^m}$ $\rightarrow$  $\mathrm{\Tilde{S_{t,n}^m}}$ $\rightarrow$   $\mathrm{\Tilde{X_{t,n} ^m}}$ $\rightarrow$   $\mathrm{\Tilde{Z_{t,n}^m}}$ are imbued in \eqref{LearningDictionaries1}.

\begin{figure}[t!]
\begin{center}
\begin{minipage}[b]{0.45\linewidth}
  \centering
  \scriptsize
  \centerline{\includegraphics[height=3.0cm]{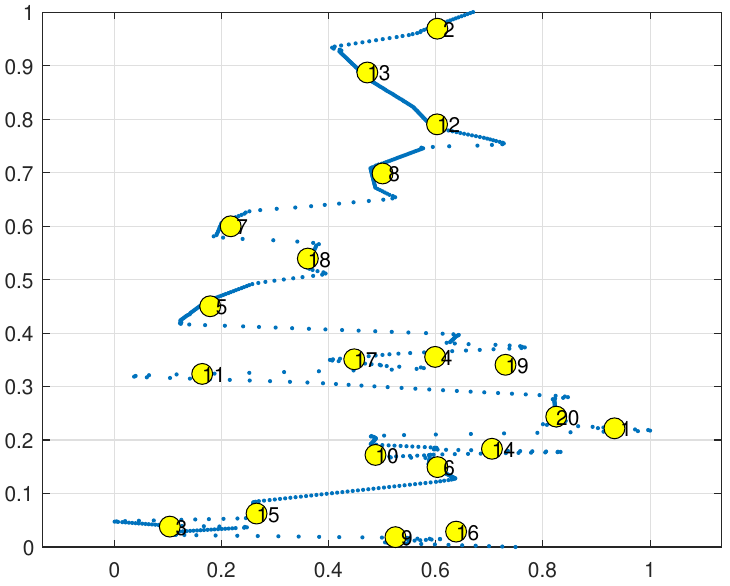}}
  \centerline{(a)}
\end{minipage}
\begin{minipage}[b]{0.45\linewidth}
  \centering
  \scriptsize
\centerline{\includegraphics[height=3.0cm]{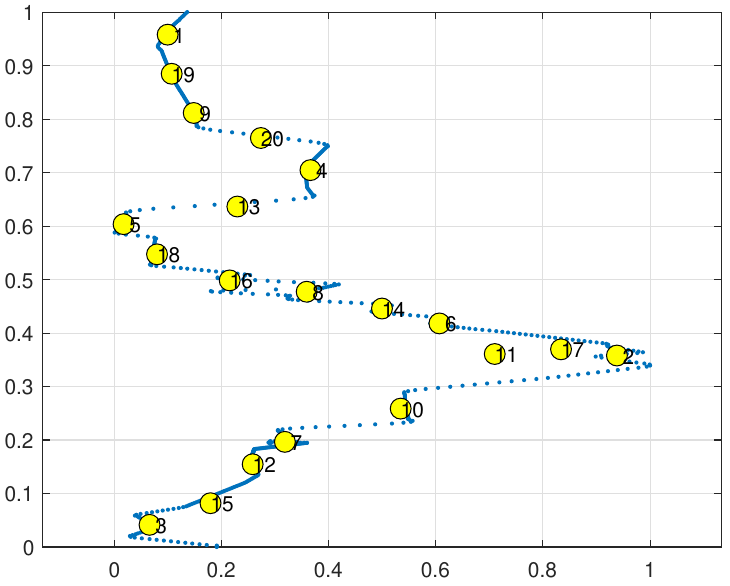}}
\centerline{(b)}
\end{minipage}
\caption{Clustering trajectories using GNG: (a) Vehicle $4$, (b) Vehicle $10$.}
\label{fig_Clustering}
\end{center}
\end{figure}

The cognitive BS instantiates the perception of the vehicular cell network with infantile knowledge (based on static assumptions regarding environmental state evolution). Thus the BS predicts the vehicle trajectory or activated V2V links at each vehicle by enacting a Null-force filter based on the model: 
$\mathrm{\Tilde{X}_{t,n}^m}$ = $\mathrm{A\Tilde{X}^m_{t-1,n}}$+ $\Tilde{w}_{t,n}^{m}$.  
This model diverges from \eqref{eq_contin1Dynamic_model} due to the control vector being null ($\mathrm{\Tilde{U}_{{S}_{t,n}}^{m}} = 0$). 
Hence, to explore the dynamic rules informing the environmental states over time, new rules need to be ascertained by exploiting generalized errors (GE) which is the difference between the observation and prediction. The GEs projected at the different hierarchical level are as follows: for the measurement space, $\Tilde{\varepsilon}_{Z_t}^{m}$ = $\mathrm{\Tilde{Z}}_{t,n}^{m}$ - $\mathrm{H \Tilde{X_{t,n}^{m}}}$ and for the generalized state space,
%
%
$\Tilde{\varepsilon}_{{\mathrm{X}_{t,n}^{m}}} = \mathrm{H}^{-1}\mathrm{Z}_{t,n}^{m} - \mathrm{X}_{t,n}^{m}$.
Growing Neural Gas (GNG), an unsupervised clustering algorithm is employed on the GEs  
and ensures dimensionality reduction of the environment representational data as illustrated  Fig.~\ref{fig_Clustering} into distinct identifiable symbols which enable our approach to computationally tractable and intrepretable. Assembling symbols to form letters which engender Words have dynamic and vivid representational properties of the global state space.  Furthermore, Fig.~\ref{fig_Interaction_Matrix} illustrates the knowledge encoded about the dynamics at different scales of this hierarchical representation. Clusters encoding the discrete variables are symbolised as 
$\mathrm{\tilde{S}_{t}^{m}}$ = $({\mathrm{\tilde{S}_{1}^{m}}}$, ${\mathrm{\tilde{S}_{2}^{m}}}$, ${\mathrm{\tilde{S}_{3}^{m}}}$, \dots, ${\mathrm{\tilde{S}_{{\mathcal{L}_{n}}}^{m}}})$,  where $\mathcal{L}_{n}^{m}$ denotes the number of clusters (aka letters) of the $m$ modality signal type at $n^{th}$ vehicle and each cluster $\mathrm{\tilde{S}_{{\mathcal{L}}_{n}^{m}}} \in $
$\mathrm{\tilde{S}^{m}}$ follows a Gaussian model characterised by $\mathrm{\tilde{S}_{{\mathcal{L}_n}}^{m}} \sim \mathcal{N}(\tilde{\mu}_{\mathrm{\tilde{S}}_{\mathcal{L}_{n}^{m}}} =  [\mu_{{\mathrm{\tilde{S}}_{\mathcal{L}_{n}^{m}}}}, 
{\dot{\mu}}_{{\mathrm{\tilde{S}}_{\mathcal{L}_n}^{m}}}], \Sigma_{\mathrm{\Tilde{S}}_{{\mathcal{L}}^{m}}})$. 
Information on the dynamic state transitions (between letters) are encoded into a time varying transition matrix $(\Pi_{\tau})$ by learning the time mapped transition probabilities $\pi_{i,j} = P(\mathrm{\tilde{S}}_{t}^{m} = i | \mathrm{\tilde{S}}_{t-1}^{m} = j, \tau)$ where $\tau$ refers to the time spent in cluster $j$ before shifting on to cluster $i$. An additional sub-dictionary of letters $\mathcal{D}_{l}^{m}$ for $m= c$ is formed to accommodate for the letter state transition mapping and future network state tracking at each individual vehicle in the V2X network.

\subsection{Learning Coupled Multi-GDBN}
\begin{figure}[t!]
    \centering
    \begin{minipage}[b]{0.49\linewidth}
      \centering
      \scriptsize
      \centerline{\includegraphics[height=2.9cm]{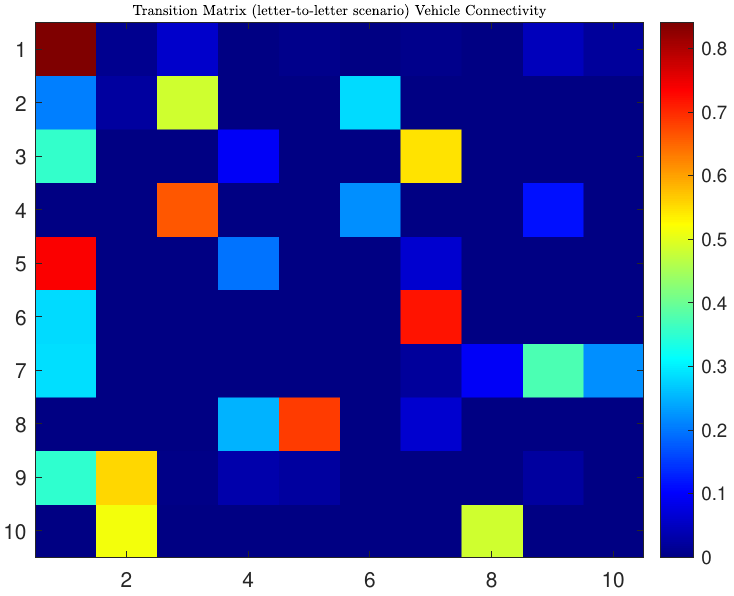}}
      \centerline{(a)}
    \end{minipage}
    \begin{minipage}[b]{0.49\linewidth}
      \centering
      \scriptsize
    \centerline{\includegraphics[height=2.9cm]{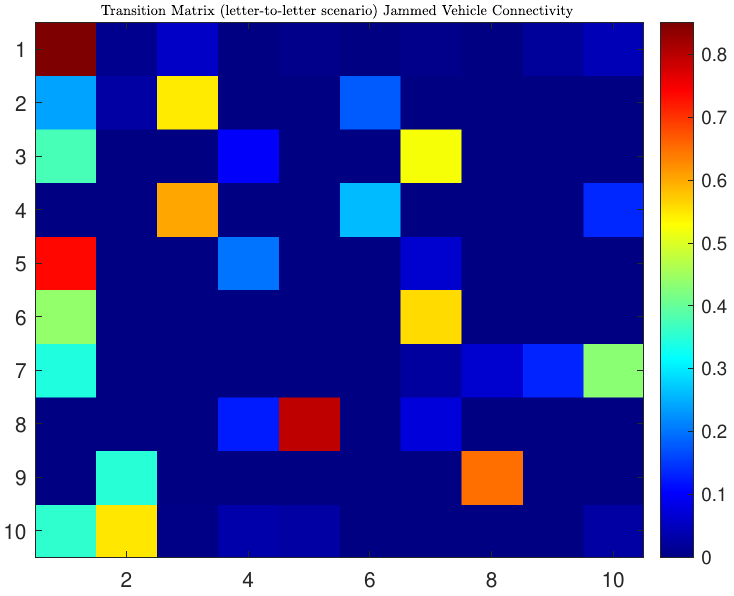}}
    \centerline{(b)}
    \end{minipage}
\\[1.5mm]
    \begin{minipage}[b]{0.49\linewidth}
      \centering
      \scriptsize
\centerline{\includegraphics[height=2.9cm]{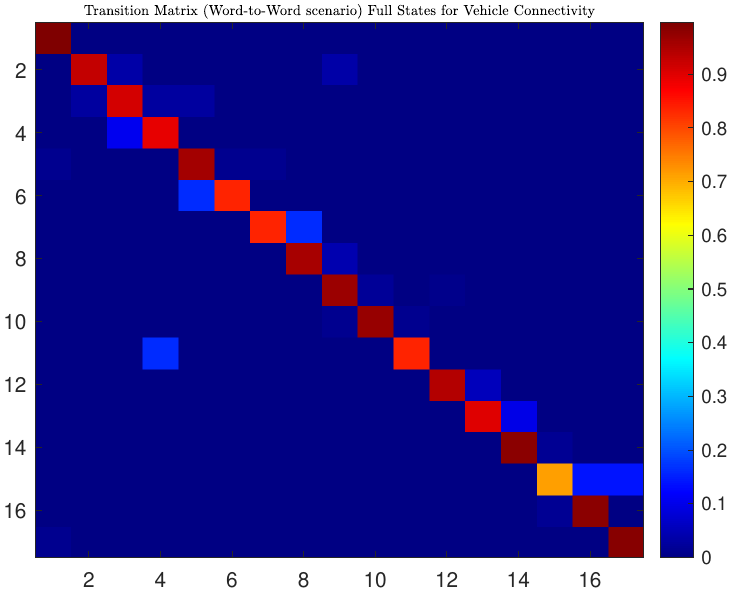}}
    \centerline{(c)}
    \end{minipage}
    \begin{minipage}[b]{0.49\linewidth}
      \centering
      \scriptsize
    \centerline{\includegraphics[height=2.9cm]{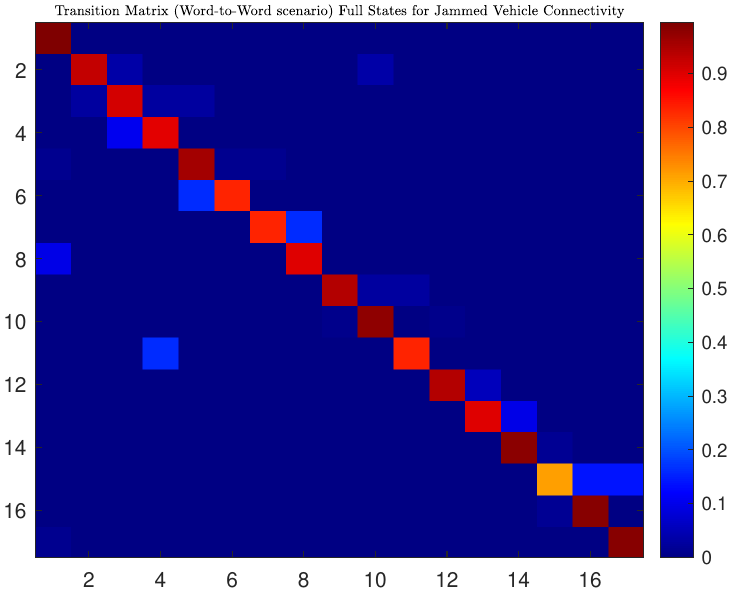}}
    \centerline{(d)}
    \end{minipage}
\\[1.5mm]
    \begin{minipage}[b]{0.49\linewidth}
      \centering
      \scriptsize
    \centerline{\includegraphics[height=3.3cm]{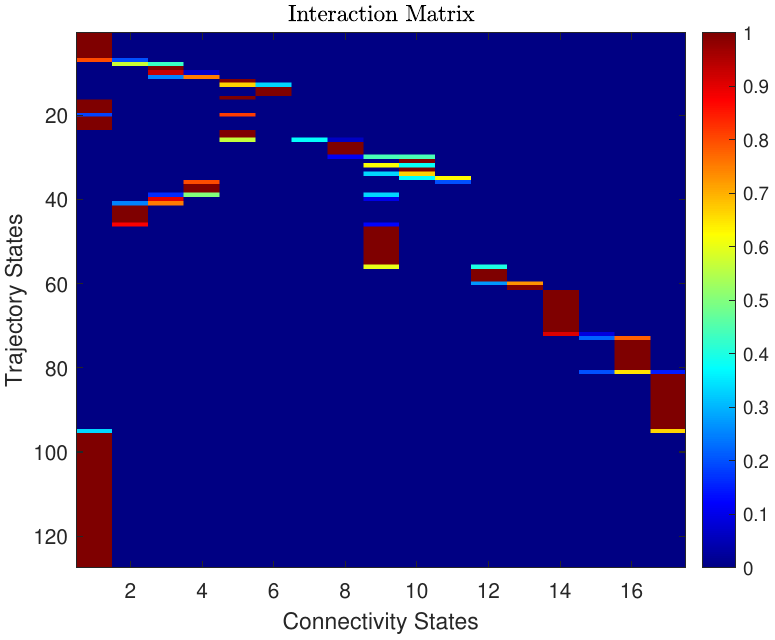}}
    \centerline{(e)}
    \end{minipage}
\caption{Learning stage: Letter-TM for (a) Normal and (b) Jammer. Word-TM for (c) Normal and (d) Jammer and (e) Interaction Matrix for N = $4$ scenario.}
\label{fig_Interaction_Matrix}
\end{figure}

The dictionary learning stage mentioned in the prior section is effectuated by learning a letter subdictionary and the symbolized letters transition probability model. Letters from subdictionaries of N vehicles are time synchronised to form a global Word at every time instance $t$. Each global Word here represents a configuration of a set of N vehicles on the road and the corresponding active V2V network connectivity graph between the N vehicles. Post fusion of Multi-GDBNs for each module $m$, we analyse time synchronized joint multi-cluster interaction behaviour between vehicle trajectory configurations and the respective Global V2X network which contains the V2V and V2I links in this scenario. By tracking the interaction, we imply the identification of multi-cluster activation across both modalities  during a viable experience. This learning and interaction is embodied into a Coupled Multi-GDBN (CM-GDBN) composed of N GDBN representing each signal modality (Trajectory and activated V2V connections). Thus, a stochastic coupling of the activated hidden states across the Multi-GDBNs in the two modalities $(\mathrm{\tilde{C}}_{t})\text{=} [\mathrm{\tilde{W}}_{t}^{p}, \mathrm{\tilde{W}}_{t}^{c}]$ at time $t$ is performed. An Interactive Matrix $\Phi \in \mathbb{R}^{W^p,W^c}$ is generated encoding the activated Word pairs which map into the firing clusters at the letter-level (refer to Fig.~\ref{fig_Interaction_Matrix}).
$\Phi$ allows us to infer the activated V2V links at a given time $t$ from the trajectory states (positions) of $N$ vehicles, 

\begin{equation}\label{eq_interaction_matrix}
    \scriptstyle\boldsymbol{\Phi} =     
    \begin{bmatrix} 
    \scriptstyle
    \mathrm{P}(\mathrm{\tilde{W}}_{1}^{c}|\mathrm{\tilde{W}}_{1}^{p}) & \scriptstyle\mathrm{P}(\mathrm{\tilde{W}}_{2}^{c}|\mathrm{\tilde{W}}_{1}^{p}) & \dots & \scriptstyle\mathrm{P}(\mathrm{\tilde{W}}_{{L{(W^{c})}}}|\mathrm{\tilde{W}}_{1}^{p}) \\
    \scriptstyle\mathrm{P}(\mathrm{\tilde{W}}_{1}^{c}|\mathrm{\tilde{W}}_{2}^{p}) & \scriptstyle\mathrm{P}(\mathrm{\tilde{W}}_{2}^{c}|\mathrm{\tilde{W}}_{2}^{p}) & \dots & \scriptstyle\mathrm{P}(\mathrm{\tilde{W}}_{{L{(W^{c})}}}|\mathrm{\tilde{W}}_{2}^{p}) \\
    %
    \vdots & \vdots & \ddots & \vdots \\
    \scriptstyle\mathrm{P}(\mathrm{\tilde{W}}_{1}^{c}|\mathrm{\tilde{W}}_{1}^{p}) & \scriptstyle\mathrm{P}(\mathrm{\tilde{W}}_{2}^{c}|\mathrm{\tilde{W}}_{{L{(W^{p})}}}) & \dots & \scriptstyle\mathrm{P}(\mathrm{\tilde{W}}_{{L{(W^{c})}}}|\mathrm{\tilde{W}}_{{L{(W^{p})}}})
    \end{bmatrix}    
\end{equation}
 where ${L}{(W^{{p} \text{ or } {c}})}$ represents the total number of words in the Dictionary of each modality $m$. 
 

\subsection{Joint Trajectory-Network Inference and Perception}
To effectuate cognitive tasks and decisions, a Bayesian generative filter named Interactive Modified Markov Jump filter (IM-MJPF) illustrated in Fig.~\ref{fig_InterCoupled_MultiGDBN} in detail is employed to access the knowledge embedded in CM-GDBN. The BS starts inferring the network connectivity map of the V2X network based on reported vehicular node positions in the cell using IM-MJPF. IM-MJPF is based on the Modified Markov Jump filter (M-MJPF) in \cite{9858012} howbeit with the introduction of a bank of GDBNs (multi-GDBNs) at each informed modalities. It is important to recollect that the M-MJPF hosts a Particle filter (PF) and Kalman filter (KF) at the discrete and continuous level, respectively. Thus aiding in multiple predictions that propagate in a feed-forward (top-down) and feedback (bottom-up) manner which modify and enforce inherent beliefs about the state of the vehicular configurations and corresponding on road active V2V communication links. The PF contained in IM-MJPF proliferates a set of (P) equally weighted $(1/L)$ particles, such that $\mathrm{\Tilde{W}}_{t}^{p}, \mathrm{\tilde{S}}_{t,n}^{p}$  $\sim {{\pi({\tilde{W}_{t,n}^p})}}$, where  ${\tilde{W}_{t,n}^p}$ positional state of the global scene and $L$ being number of vehicles on the road.
The BS uses $\Phi$ in \eqref{eq_interaction_matrix} to infer $({\mathrm{\tilde{W}}_{t}^c})$ the vehicular network map at time $t$ and by further deduction into discrete superstate variables $\mathrm{\tilde{S}}_{t,n}^{p}$. 
A PF is used to predict future multiple superstate guided by top-down and inter-time predictions as mention in \eqref{eq_discreteDynamic_model}.

%
%
The posterior probability is expressed as follows: $\pi({\mathrm{\tilde{W}}_{t}^p}, {\mathrm{\tilde{W}}_{t}^c})
{=} P({\mathrm{\tilde{W}}_{t}^c}|{\mathrm{\tilde{W}}_{t-1}^c}, {\mathrm{\tilde{W}}_{t-1}^p})$ and $\pi({\mathrm{\tilde{W}}_{t}^p})
{=} P({\mathrm{\tilde{W}}_{t}^p}|{\mathrm{\tilde{W}}_{t-1}^p}, {\mathrm{\tilde{W}}_{t-1}^c} )$ at the Metacluster (Word) level.
The diagnostic (bottom-up update) message in the network leading up to the meta-cluster level is as follows:
%
%
%
\begin{multline}
\scriptsize
    \lambda\left({\mathrm{\tilde{W}}}_{t}^c\right) = \prod_{n=1}^{L} \lambda\left(\mathrm{\tilde{S}}_{t,n}^{p}\right) \prod_{n=1}^{L} P\left(\mathrm{\tilde{S}}_{t,n}^{p} \mid {\mathrm{\tilde{W}}}_{t,n}^{p}\right).
\end{multline}
At the superstate discrete level the feed-forward and feed-backward messages are expressed as follows:
\begin{gather*}
 \pi\left(\mathrm{\tilde{S}}_{t,n}^{p}\right)=P\left(\mathrm{\tilde{S}}_{t,n}^{p} \mid \mathrm{\tilde{S}}_{t-1,n}^{p}, \ldots, \mathrm{\tilde{S}}_{t-1,L}^{p}\right),  L \neq n \\ 
 \text{and }
 \lambda\left(\mathrm{\tilde{S}}_{t,n}^{p}\right)=\lambda\left(\mathrm{\tilde{X}}_{t,n}^{p}\right) P\left(\mathrm{\tilde{X}}_{t,n}^{p} \mid \mathrm{\tilde{S}}_{t,n}^{p}\right).
 \end{gather*}
Posterior probabilities at the continuous state level $\tilde{X}$ mapped at individual vehicles is defined by the expression: 
\begin{multline}
\pi\left(\mathrm{\tilde{X}}_{t,n}^{p}\right) = P\left(\mathrm{\tilde{X}}_{t,n}^{p}, \mathrm{\tilde{S}}_{t,n}^{p} \mid \mathrm{\tilde{Z}}_{t-1,n}^{p}\right)
  \\ = \int P\left(X_{t,n}^{p} \mid X_{t-1,n}^{p}, S_{t,n}^{p}\right) \lambda\left(X_{t-1,n}^{p}\right) d X_{t-1,n}^{p}
\end{multline}
and the diagnostic message is,
\begin{multline}
\scriptsize
\lambda\left(\mathrm{\tilde{X}}_{t,n}^{p}\right) = P\left(\mathrm{\tilde{Z}}_{t,n}^{p}\mid \mathrm{\tilde{X}}_{t,n}^{p}\right) \sim N\left(\mu_{\mathrm{Z}_{n}^p}, \Sigma_{{\mathrm{Z}}_{n}^{p}})\right).
\end{multline}

The posterior distribution is updated by the new sensory evidence $\mathrm{\tilde{Z}_t^{m}}$ by purposing the diagnostic messages  discussed \textit{i.e.} $\lambda\left(\mathrm{\tilde{X}}_{t,n}^{m}\right)$ , $\lambda\left(\mathrm{\tilde{S}}_{t,n}^{m}\right)$ and $\lambda\left({\mathrm{\tilde{W}}}_{t}^{m}\right)$ at the State, Superstate and Word level respectively.
 
\subsection{Vehicular Network Abnormality Detection}
The BS surveys the current state of the vehicular network by verifying if the V2V communications are under attack or in the optimal state based on predicted road and network configurations.
In this work we introduce a new abnormality (Bayesian surprise) indicator at the Word level that calculates the similarity between the predicted word and the observed word encoding the $m$ modules (recall $m = p$ or $c$) based on the symmetric Kullback-Leibler-Divergence Abnormality (KLDA) \cite{9322583} defined as:
\begin{multline}
\small
{\Upsilon_{\tilde{W}_{t}^{m}}} = D_{K L}\left(\pi\left(\tilde{W}_{t}^{m}\right) \| \lambda\left(\tilde{W}_{t}^{m}\right)\right) + \\ D_{K L}\left(\lambda\left(\tilde{W}_{t}^{m}\right) \| \pi\left(\tilde{W}_{t}^{m}\right)\right).
\end{multline}
KLDA is used to compare the predictive message $\pi(\tilde{W}_{t}^{m})$ and diagnostic message $\lambda(\tilde{W}_{t}^{m})$ defined in the previous sub-section which tracks abnormalities at the global V2X network.
BS enacts to verify if the current situation is normal or not \textit{i.e.} the V2V links are under attack by a jammer based on the hypothesis test:
\begin{equation} \label{eq_WordKLDA}
\left\{\begin{array}{l}
\mathcal{H}_0: \text{if } {\Upsilon_{\tilde{W}_{t}^{m}}} < \xi  \\
\mathcal{H}_1: \text{if } {\Upsilon_{\tilde{W}_{t}^{m}}} > \xi  \\
\end{array}\right.
\end{equation}
where $\xi$ = $\mathbb{E}[{\Upsilon_{\tilde{W}_{t}^{m}}}] + \phi\sqrt{\mathbb{V}[{\Upsilon_{\tilde{W}_{t}^{m}}}]}$~\textit{(ideally $\phi = 3$)}. 
Within $\xi$, ${\Upsilon_{\tilde{W}_{t}^{m}}}$ refers to abnormality signals during training (\textit{i.e.} under reference situation with no jammer) (demonstrated as blue line in Fig. \ref{fig_KLD_ABN}). 

\subsection{Evaluation metrics}
ROC (Receiver Operating Characteristic) curves are used in this study to quantitatively assess the performance of our jamming detection scheme. ROC curves here illustrates the trade-off between the true positive rate (sensitivity) and the false positive rate (1 - specificity) for different detection thresholds. 

\section{Simulation and Experimental Results}
\begin{figure}[t!]
\begin{center}
\begin{minipage}[b]{0.45\linewidth}
  \centering
  \scriptsize
\centerline{\includegraphics[height=2.9cm]{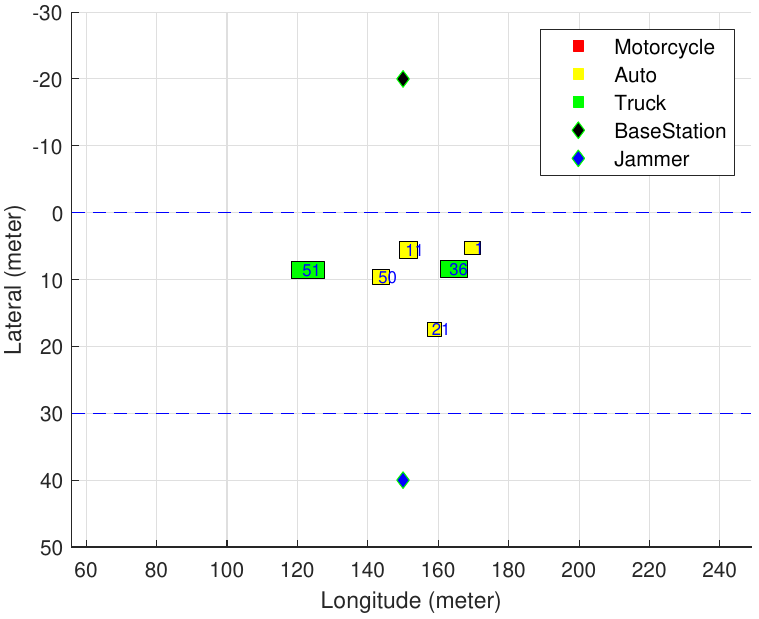}}
\centerline{(a)}
\end{minipage}
\begin{minipage}[b]{0.45\linewidth}
  \centering
  \scriptsize
\centerline{\includegraphics[height=2.9cm]{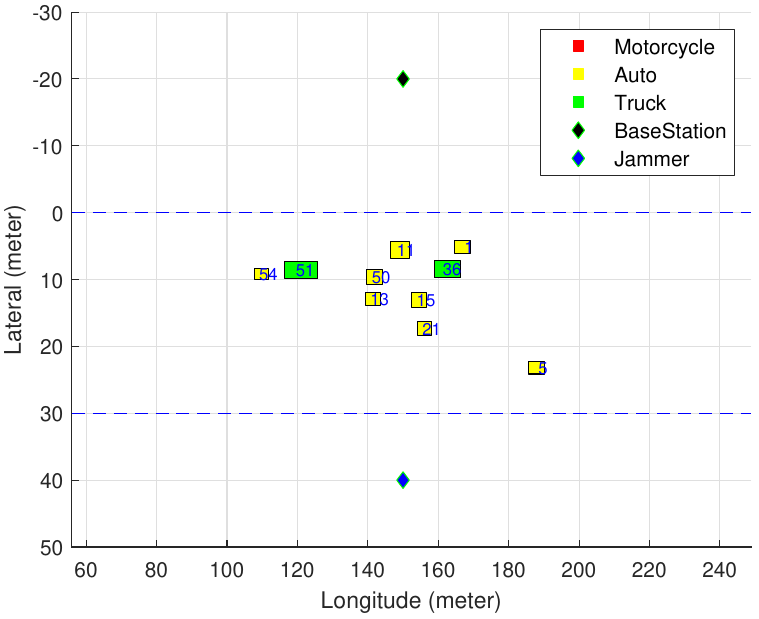}}
\centerline{(b)}
\end{minipage}
\\[1.5mm]
\begin{minipage}[b]{0.45\linewidth}
  \centering
  \scriptsize
  \centerline{\includegraphics[height=2.9cm]{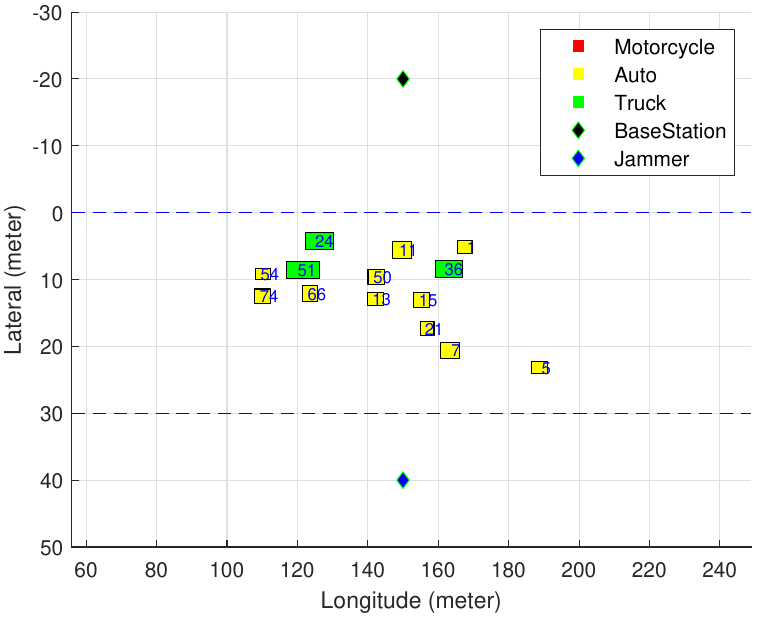}}
  \centerline{(c)}
\end{minipage}
\caption{Examples of four different sets of vehicle configuration: a) N=6, b) N=10, c) N=14  from NGSIM dataset.}
\label{fig_Veh_trajectory_clustering}
\end{center}
\end{figure}

In this section we perform a performance test of our scheme to detect vehicular network abnormalities using extensive simulations. Scenarios on the freeway with multiple number of vehicles (where N = 4, 6, 10 and 14) are studied using our detection scheme. We consider N vehicles maneuvering safely in an interactive environment relaying their positions, CSI and active V2V connections to BS. The vehicles  trace a predefined trajectory with velocities from the \textit{Interstate 80 Freeway Dataset}~\cite{ngsim}  illustrated in Fig. \ref{fig_Veh_trajectory_clustering}. performing different maneuvers on a freeway. The BS assigns one PRB to each vehicle in the cell to realize the V2I link.
In this scenario, the transmitted signal carrying the vehicle's status information and the jamming signal both utilize Quadrature Phase Shift Keying (QPSK) modulation. The simulation is configured with specific parameters: a carrier frequency of $2$ GHz, a bandwidth (BW) of $1.4$ MHz, a cell radius of $500$ m, an RSU antenna height of $25$ m with a gain of $8$ dBi, a receiver noise figure of $5$ dB, a vehicle antenna height of $1.5$ m with a gain of $3$ dBi, a V2I transmit power of $23$ dBm and jammer transmit power of $23$ dBm. Additionally, the simulation incorporates a Signal-to-Noise Ratio (SNR) of $20$ dB, a path loss model based on $(128.1+37.6\log{d})$, log-normal shadowing with a standard deviation of $8$ dB, and a fast fading channel following the Rayleigh distribution \cite{10118852}. While the standard minimum distance at which a vehicle-to-everything (V2X) nodes can detect a signal depends on the specific technology and implementation being used.  In our system model, without loss of generality with respect to short-range or cellular communications, for V2V transmit power of upto $23$ dBm is converted to a vehicle to vehicle relative distance oriented relationship. Therefore, we assume the adjacency matrix of graph $\mathcal{G}_{t}$ encodes the V2V connectivity which again is explicitly mapped to the relative distances between vehicles ${d_{ij}}$, where $i$ and $j$ indicate vehicle index number $\in$ N. The maximum connectivity distance between $i^{th}$ vehicle to $j^{th}$ vehicle for all N vehicles on the freeway is defined by scalar value $d_k$ (we assume $d_k$ $\le 10m$ as connected) and thereby transforming by abstraction the relative distance matrix to connectivity adjacency matrix of the global vehicular network. 

\begin{figure}[t!]
\centering
    \begin{minipage}[b]{0.52\linewidth}
      \centering
      \scriptsize
      \centerline{\includegraphics[height=2.5cm]{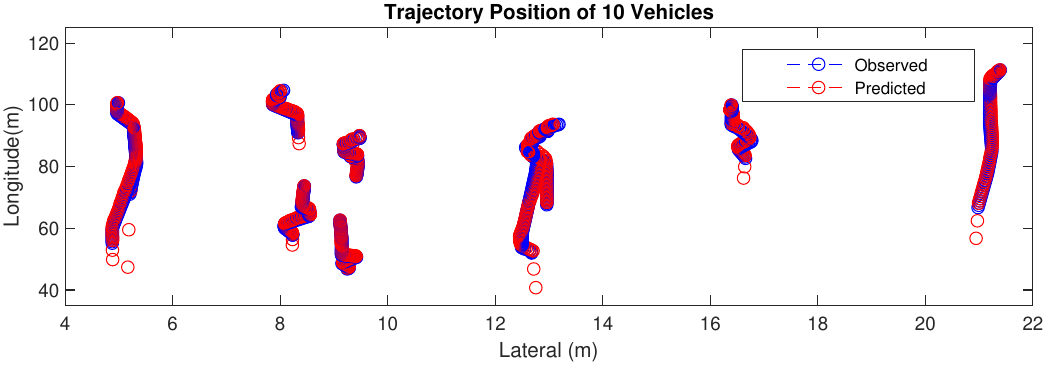}}
      \centerline{(a)}
    \end{minipage}
    \\[1mm]
    \begin{minipage}[b]{0.52\linewidth}
      \centering
      \scriptsize
    \centerline{\includegraphics[height=2.5cm]{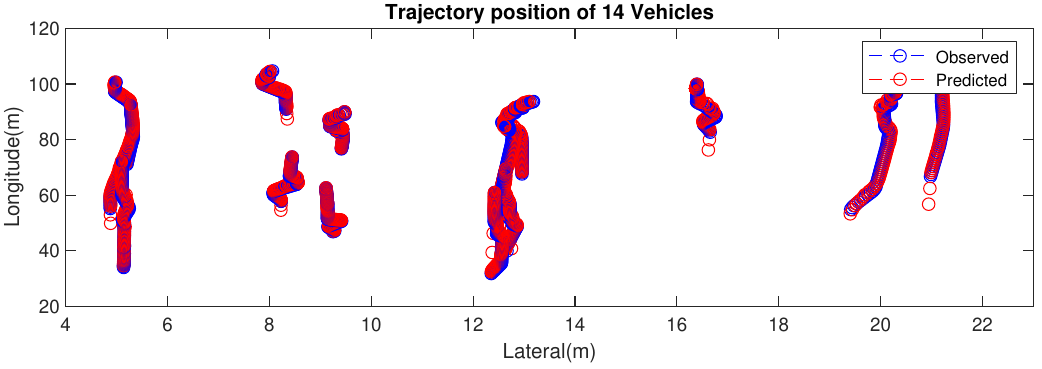}}
    \centerline{(b)}
    \end{minipage}
\caption{Predicted Trajectories at the BS for different number of Vehicles: (a) N = 10, (b) N = 14.}
\label{fig_PredTraj}
\end{figure}

\begin{figure}[t!]
\begin{center}
\begin{minipage}[b]{0.49\linewidth}
  \centering
  \scriptsize
  \centerline{\includegraphics[height=3.4cm]{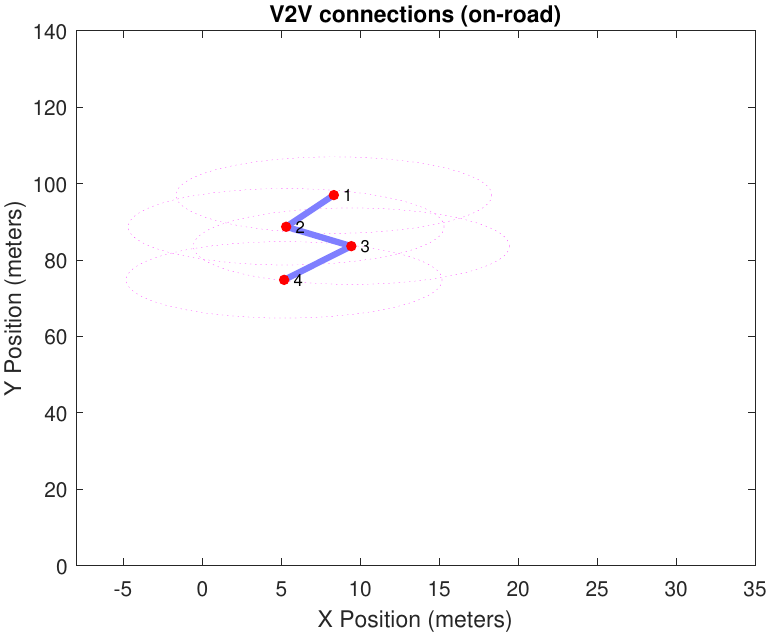}}
  \centerline{(a)}
\end{minipage}
\begin{minipage}[b]{0.49\linewidth}
  \centering
  \scriptsize
\centerline{\includegraphics[height=3.4cm]{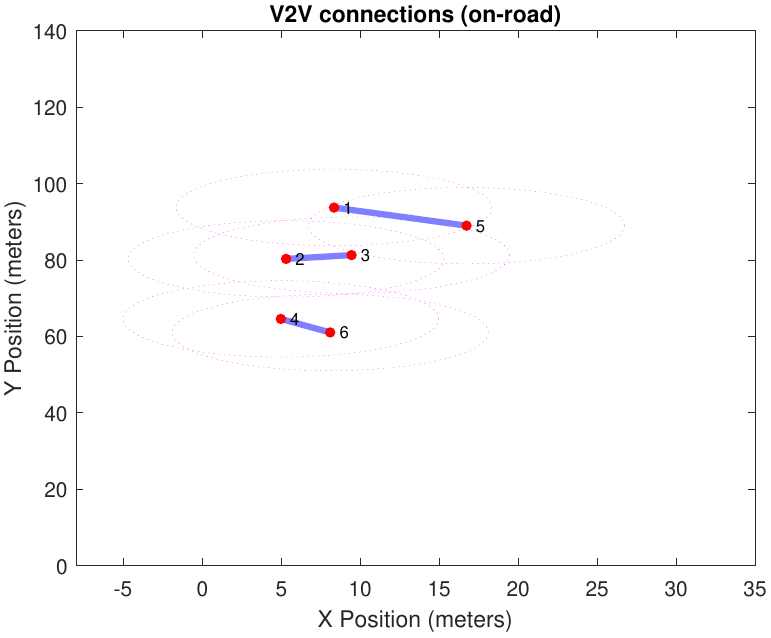}}
\centerline{(b)}
\end{minipage}
\\[1.5mm]
\begin{minipage}[b]{0.49\linewidth}
  \centering
  \scriptsize
\centerline{\includegraphics[height=3.4cm]{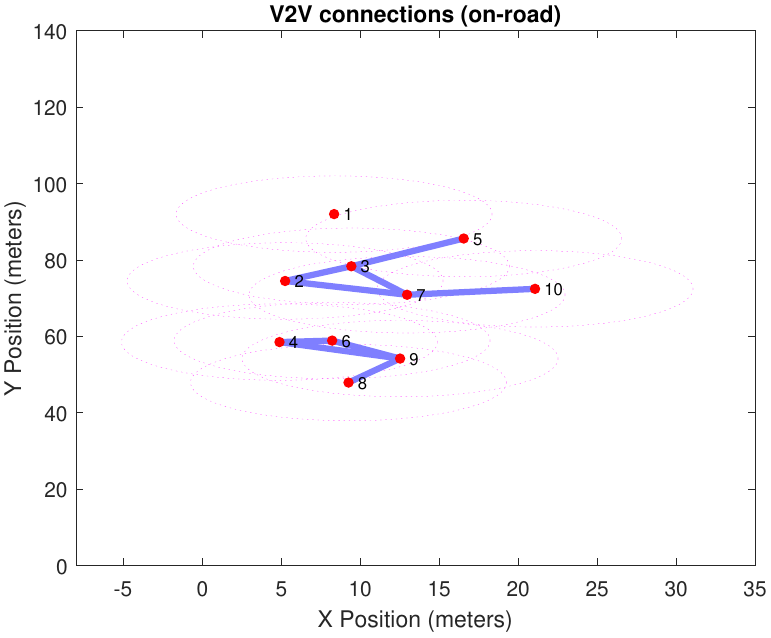}}
\centerline{(c)}
\end{minipage}
\begin{minipage}[b]{0.49\linewidth}
  \centering
  \scriptsize
\centerline{\includegraphics[height=3.4cm]{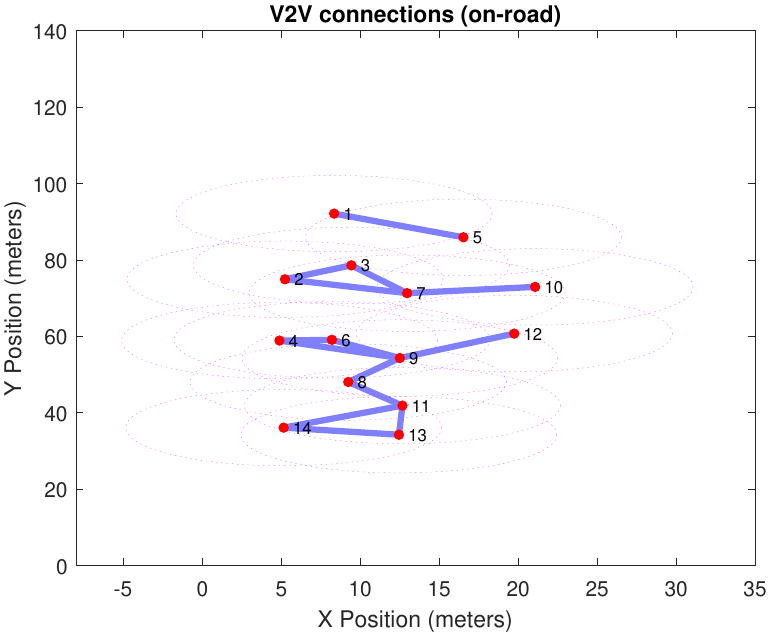}}
\centerline{(d)}
\end{minipage}
\caption{Predicted Network Graphs $\mathcal(G_t)$ using Dictionaries: (a) $4$ Vehicles, (b) $6$ Vehicles, (c) $10$ Vehicles, and (d) $14$ Vehicles.}
\label{fig_vehicular_nw_pred}
\end{center}
\end{figure}

The multi-vehicle tracking capability of the BS is demonstrated via trajectory prediction in Fig.~\ref{fig_PredTraj} by inferring from variables that encode spatio-temporal relationships between on road vehicles embedded inside CM-GDBN during the learning phase.
The spatio-temporal configurations of vehicles affect the overlaying dynamic V2V connections persistent during an interaction/manoeuvre experience. Consequently, the connectivity map between vehicles can be estimated based on the spatio-temporal vehicular interactions as depicted in Fig.~\ref{fig_vehicular_nw_pred}. 

\begin{figure}[t!]
\begin{center}
\begin{minipage}[b]{0.49\linewidth}
  \centering
  \scriptsize
\centerline{\includegraphics[height=3.4cm]{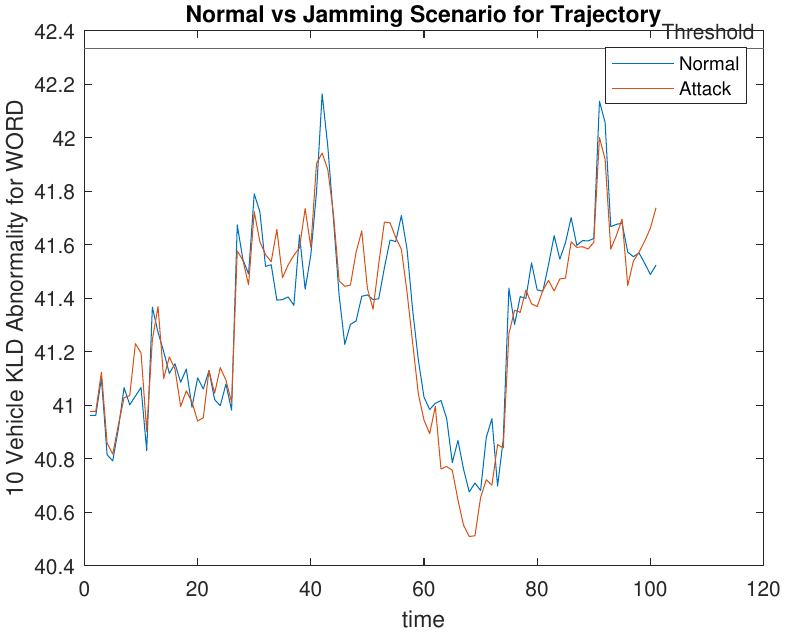}}
  \centerline{(a)}
\end{minipage}
\begin{minipage}[b]{0.49\linewidth}
  \centering
  \scriptsize
\centerline{\includegraphics[height=3.4cm]{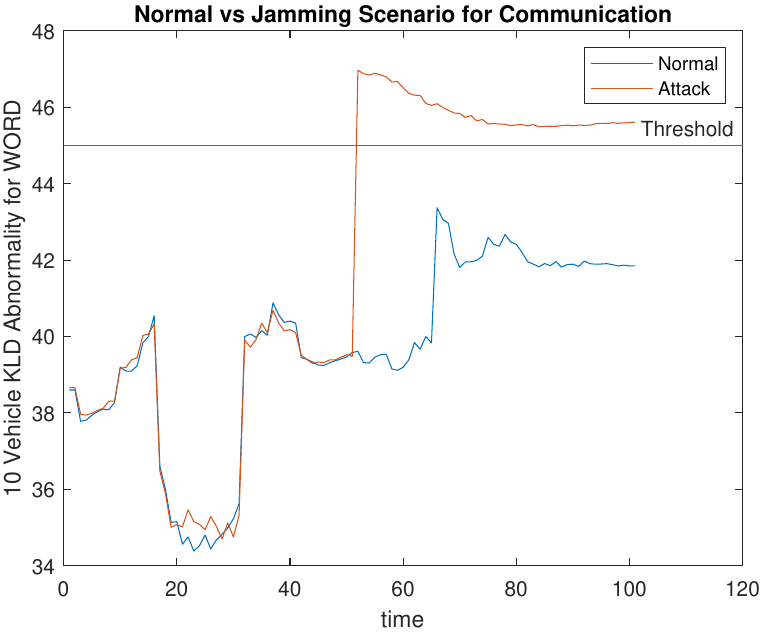}}
\centerline{(b)}
\end{minipage}
\\[1.5mm]
\begin{minipage}[b]{0.49\linewidth}
  \centering
  \scriptsize
\centerline{\includegraphics[height=3.4cm]{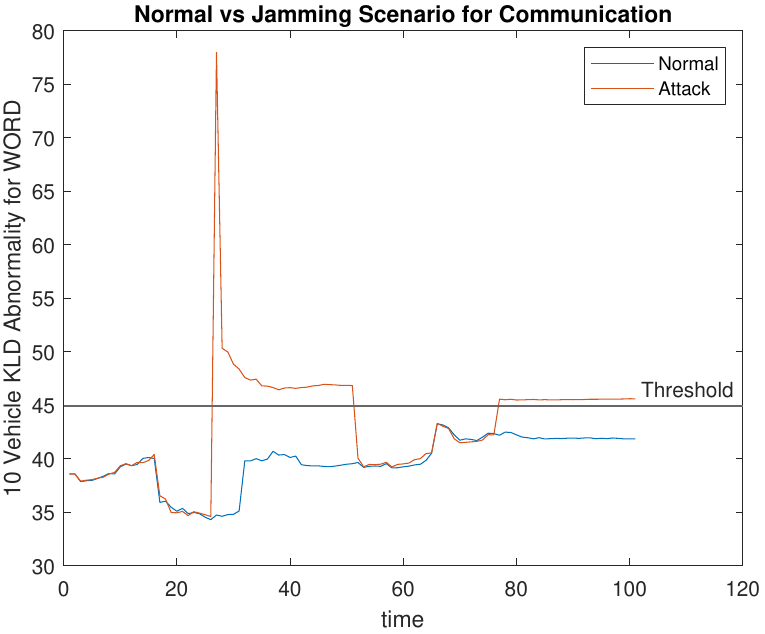}}
\centerline{(c)}
\end{minipage}
\caption{BS tracking abnormality: (a) Trajectory Normality, (b) Abnormality (single window attack), c) Abnormality (double window attack) at the Network (Word) level for N = 10 scenario.}
\label{fig_KLD_ABN}
\end{center}
\end{figure}

\begin{figure}[t!]
    \centering
    \begin{minipage}[b]{0.49\linewidth}
      \centering
      \scriptsize
      \centerline{\includegraphics[height=3.4cm]{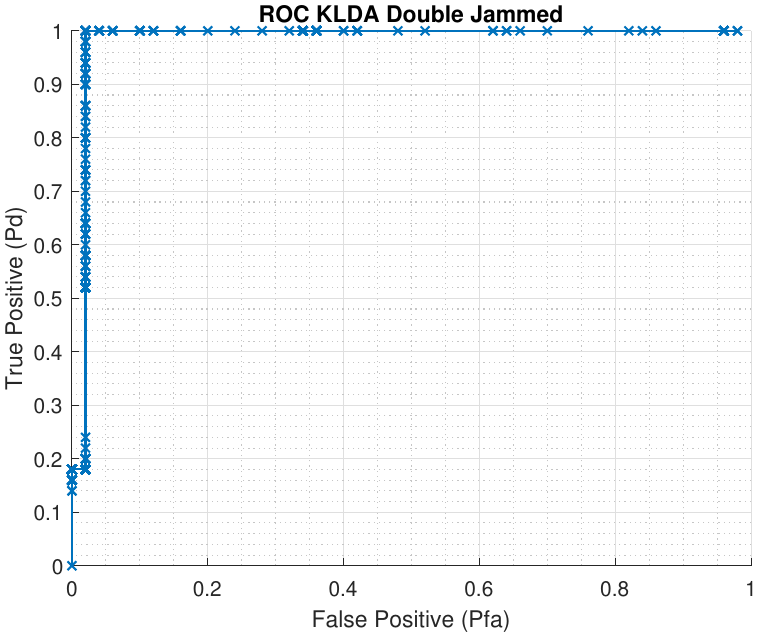}}
      \centerline{(a)}
    \end{minipage}
    \begin{minipage}[b]{0.49\linewidth}
      \centering
      \scriptsize
    \centerline{\includegraphics[height=3.4cm]{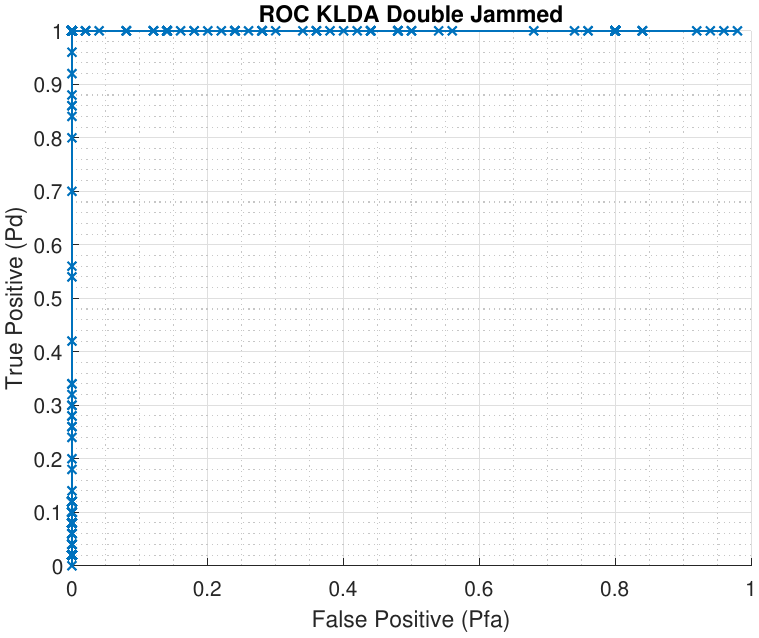}}
    \centerline{(b)}
    \end{minipage}
\\[1.5mm]
    \begin{minipage}[b]{0.49\linewidth}
      \centering
      \scriptsize
    \centerline{\includegraphics[height=3.4cm]{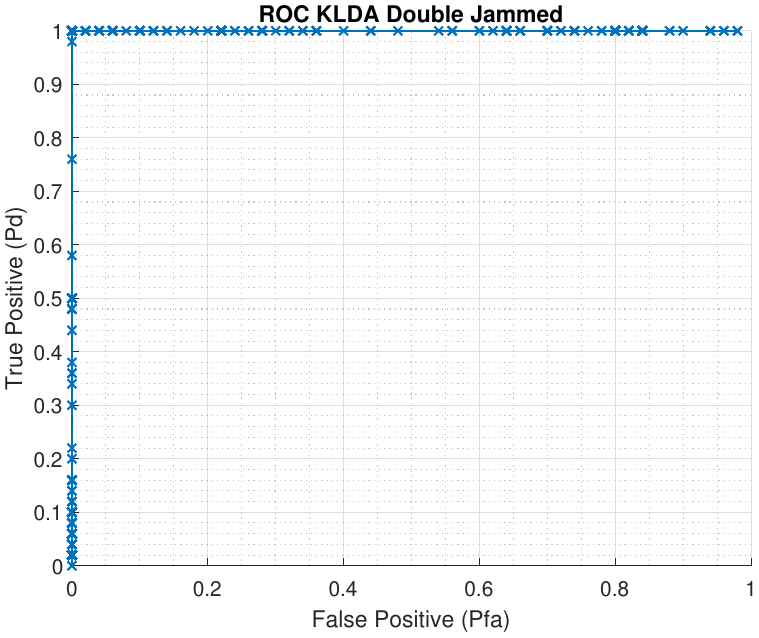}}
    \centerline{(c)}
    \end{minipage}
    \begin{minipage}[b]{0.49\linewidth}
      \centering
      \scriptsize
    \centerline{\includegraphics[height=3.4cm]{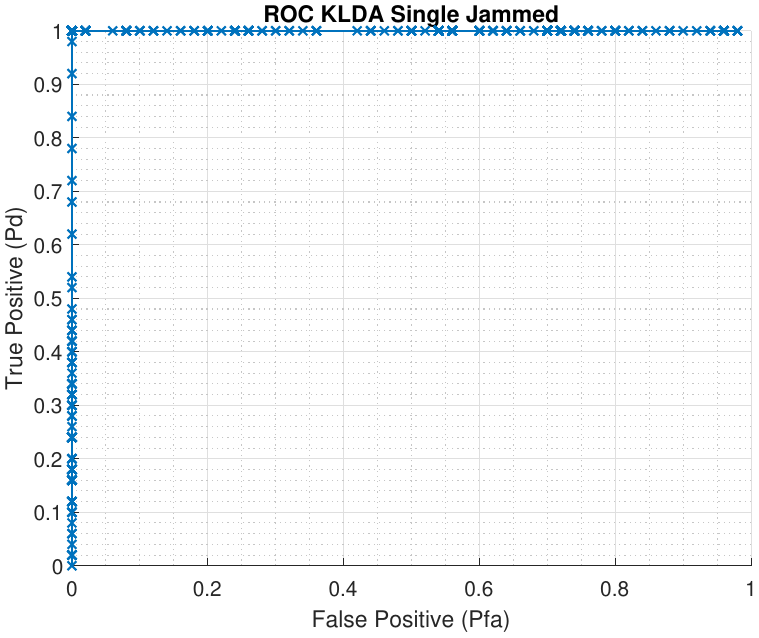}}
    \centerline{(d)}
    \end{minipage}
\caption{ROC (KLDA) Curves for Jammer detection: (a) 4 Vehicle Double-attack, (b) 6 Vehicle Double-attack, (c) 10 Vehicle Double-attack, (d) 10 Vehicle  Single-attack.}
\label{fig_ROC_Curves}
\end{figure}

The optimal decision making process at AI enabled cognitive BS is elucidated in Fig.~\ref{fig_KLD_ABN}, where mismatches between expected and observed words trigger regular abnormality significantly to cross a predefined threshold as defined in (\ref{eq_WordKLDA}). The abnormality tracking results for single and multi attack demonstrate the effectiveness of the proposed framework and engender us to interpret the observed phenomenon in more explainable fashion through causal reasoning capability  of PGMs. Moreover, in this study we performance test our approach to detecting anomalies caused by the (two) jamming attack types, single and multiple attack over different window intervals.
ROC curves verify the efficacy of our jamming detection scheme for the attack scenarios in Fig.~\ref{fig_ROC_Curves} by learning the optimum environment and thereby embedding the knowledge into an optimal dynamic Bayesian network. 
While the ROC performance seen here are strong, it may be duly noted that as multiplicity of attacks increase ROC performance remains adept at the KLDA level in comparison to random observed abnormalities.

%
%
%


%

%

\section{Conclusion}
%
Our proposed approach uses Multi-CDBNs to imbibe new observations at scale into its model over time, thus enabling incremental learning and inherent explainability. This approach performs predictions based on past observed data, which provides an interpretable representation of the model's reasoning. By predicting future V2X network states and structure through vehicle trajectories on the freeway, discrepancies in vehicular connectivity can be detected by matching the connectivity prediction with on-road radio environment observations. Our framework uses a Bayesian generative mechanism that effectuates probabilistic belief-based inferences of the current and future network connectivity state, inspired by the Bayesian Brain hypothesis. This approach has been validated successfully via multiple simulation tests. In the future, we plan to perform attack localization and sensitivity analysis tests for varying network sizes and channel conditions/impairments to evaluate and further validate our proposed framework.

\bibliographystyle{unsrt}
\bibliography{References}

\end{document}